  \documentstyle[twocolumn,prl,aps,epsfig,epsf, amssymb,amsfonts,psfrag]{revtex}
\begin{document}
%
%
\renewcommand{\Re}{\operatorname{Re}}
\renewcommand{\Im}{\operatorname{Im}}
\newcommand{\Tr}{\operatorname{Tr}}
\newcommand{\sign}{\operatorname{sign}}
\newcommand{\dd}{\text{d}}
\newcommand{\q}{\boldsymbol q}
\newcommand{\p}{\boldsymbol p}
\newcommand{\rr}{\boldsymbol r}
\newcommand{\pp}{p_v}
\newcommand{\vv}{\boldsymbol v}
\newcommand{\I}{{\rm i}}
\newcommand{\pphi}{\boldsymbol \phi}
\newcommand{\ds}{\displaystyle}
\newcommand{\be}{\begin{equation}}
\newcommand{\ee}{\end{equation}}
\newcommand{\bea}{\begin{eqnarray}}
\newcommand{\eea}{\end{eqnarray}}
\newcommand{\Acl}{{\cal A}}
\newcommand{\Rcl}{{\cal R}}
\newcommand{\Tcl}{{\cal T}}
\newcommand{\Tmin}{{T_{\rm min}}}
\newcommand{\Toff}{{\langle \delta T \rangle_{\rm off} }}
\newcommand{\Roff}{{\langle \delta R \rangle_{\rm off} }}
\newcommand{\RoffI}{{\langle \delta R_I \rangle_{\rm off} }}
\newcommand{\RoffII}{{\langle \delta R_{II} \rangle_{\rm off} }}
\newcommand{\dg}{{\langle \delta g \rangle_{\rm off} }}
\newcommand{\rd}{{\rm d}}
\newcommand{\br}{{\bf r}}
\newcommand{\la}{\langle}
\newcommand{\ra}{\rangle}

\twocolumn[\hsize\textwidth\columnwidth\hsize\csname @twocolumnfalse\endcsname
%
%
\draft

\title{Spin Filter Effects in Mesoscopic Ring Structures}

\author{Markus Popp$^{(1)}$,
 Diego Frustaglia$^{(2)}$,
        and Klaus Richter$^{(1)}$ }

\address{
$^{(1)}$ Institut f{\"u}r Theoretische Physik, Universit{\"a}t Regensburg, 
         93040 Regensburg, Germany}
\address{
$^{(2)}$ Institut f{\"u}r Theoretische Festk{\"o}rperphysik,
 Universit{\"a}t Karlsruhe, 76128 Karlsruhe, Germany }

\date{\today}

\maketitle
\begin{abstract}
We study  the spin-dependent conductance of ballistic mesoscopic ring systems in the presence of an inhomogeneous magnetic field. We show   that, for the  setup proposed, even a small Zeeman splitting can lead to a considerable spin 
 polarisation of the current. Making use of a spin-switch effect \cite{FHR01} we propose a device of two rings connected in series that  in principle allows for both creating and  coherently controlling spin polarized currents at low temperatures.
\end{abstract}

\pacs{ }    

]

\vspace{1cm}


\narrowtext
Since the proposal of the Datta-Das transistor \cite{Datta-Das} over a decade ago, much effort has been spent on finding an effective mechanism for achieving spin-polarized electron injection into semiconductors \cite{SpinI}. The ability to inject and detect spin-polarized currents in a semiconducting material widens the field of usual magneto-electronics in metals and opens up the intriguing program of performing spin electronics \cite{Prinz} based on nonmagnetic semiconductor devices. Due to the obstacle of the conductivity mismatch \cite{Schmidt}  it has so far  proved  difficult to demonstrate  polarized spin injection from a ferromagnetic metallic contact into a semiconductor. An alternative approach, which is based on magnetic semiconductors \cite{Spininj-F,Spininj-O},  gives excellent results concerning the injection efficiencies but has the drawback that it is, at least up to now, restricted to low temperatures. The mechanisms commonly used for creating spin-polarized currents rely on magnetic materials or a large Zeeman splitting in a homogeneous magnetic field \cite{comm-filt}.  In this work we  study ballistic mesoscopic rings with \emph{inhomogeneous} magnetic fields and show that even a small Zeeman splitting (compared to the Fermi energy) can lead to a considerable spin-polarisation of the current. Usually the Zeeman splitting is exploited to align the spin to the energetically favourable lower Zeeman level. Here we demonstrate that for the presented field texture  the electron spin, which occupies the \emph{higher} Zeeman level, is more likely to align itself with the local field direction and thus contributes to a larger degree to the current. The described effect is most pronounced in the adiabatic regime of strong  fields where the electron spin follows the spatially varying direction of the magnetic field. But our numerical calculations show that there also exists a region of moderate field strengths for which a spin-polarisation of about 30\% can be achieved. Such spin-injection efficiencies could be realised using  GaAs/AlGaAs heterostructures  with  low carrier densities. This  has the advantage that both the injector and a possible  spin controlling device, such as the spin-switch  \cite{FHR01}, can be fabricated from the same material and negative interface effects would be avoided.

\begin{figure}[htbp]
 \begin{minipage}[b]{0.45 \linewidth }  
  \begin{center}
    
    \psfrag{a}{(a)}
    \centering \epsfig{file=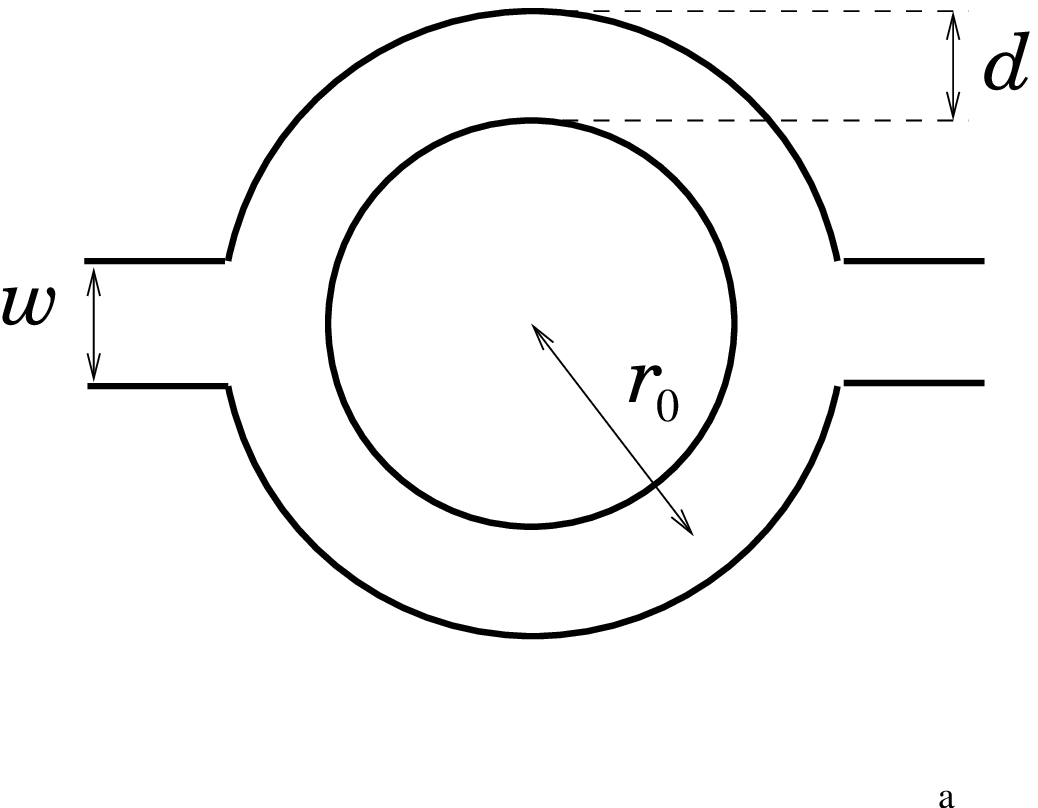, width=\linewidth} 
         
   \end{center}
 \end{minipage} \hfill
 \begin{minipage}[b]{0.45 \linewidth }  
  \begin{center}
     \psfrag{Si}{$\rm \vec S_{in} $}
    \psfrag{So}{$ \rm\vec S_{out} $}
    \psfrag{b}{(b)} 
    \centering \epsfig{file=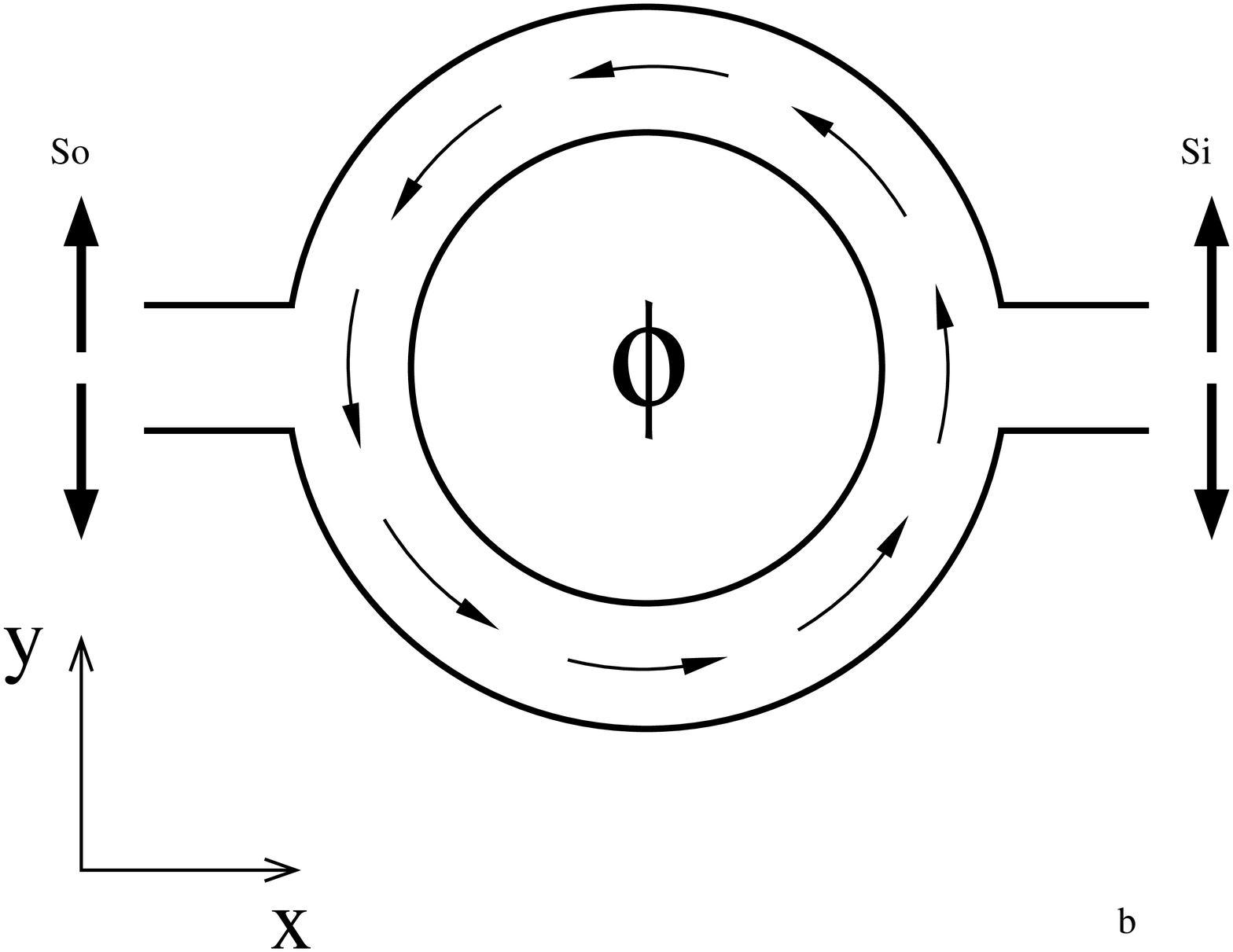, width=\linewidth}

   \end{center}
 \end{minipage}
\vspace*{3mm}
 \caption{(a) Example of a quasi 1d ring structure with mean radius $r_0$ and width $d$ as used for the numerical calculations. (b) Illustration of the ring  with circular in-plane field and flux $\Phi$ through the ring. We assume electrons coming from the right and define the spin polarisation with respect to the y-axis.}
 \label{ring} 
\end{figure}

\begin{figure}[h]
   \begin{center}

     \psfrag{b}{ }
     \psfrag{r}{$r_0$}
     \psfrag{B}{$\vec B$}

    \centering \epsfig{file=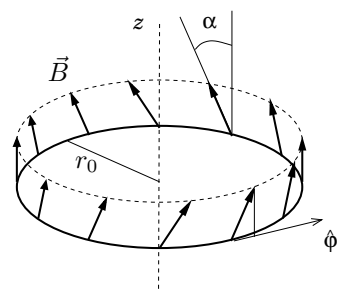, width=0.5\linewidth} 
\vspace{1mm}       
     \caption{ The total magnetic field  $\vec B(r_0)$ in the ring of Fig. \ref{ring} is composed of an inhomogeneous in-plane field $\vec B_i = B_i(r_0) \  \hat \varphi $ and a perpendicular homogeneous field  $\vec B_0 = B_0 \ \hat z$.}
     \label{field}
   \end{center}
 \end{figure}

To be more specific we consider
 symmetric 1d and 2d ballistic  mesoscopic rings  with two attached leads as shown in Fig. \ref{ring}. The chosen magnetic field texture is illustrated in Fig. \ref{field}. 

 For the inhomogeneous magnetic field we assume a circular configuration $\vec B_i (\vec r)\sim 1/r \  \hat \varphi $ (in polar coordinates) centered around the inner disk of the microstructure. Such a field can be viewed as being generated by a perpendicular electrical current through the disk \cite{circ-ring}. The magnetic flux through the ring can be varied via a homogeneous field $\vec B_0 = B_0 \ \hat z$ pointing in the z-direction.
The Hamiltonian for noninteracting electrons with effective mass $m^*$ and 
charge $-e$, in the presence of a magnetic field $\vec B(\vec r)=\nabla \times \vec A(\vec r) $, reads
\begin{equation}\label{H}
H=\frac{1}{2m^*}\left[\vec p + \frac{e}{c} 
\vec A(\vec r) \right]^2 + V(\vec r) + \mu\vec B(\vec r) \cdot\vec \sigma. 
\end{equation}
The coupling of the electron spin to the magnetic field enters via the Zeeman term $\mu\vec B(\vec r) \cdot\vec \sigma$, where $\vec \sigma$ is the 
Pauli spin vector and $\mu = g^*e\hbar/(4m_0 c)$ the magnetic moment with 
$g^\ast$ the gyromagnetic ratio. 
The electrostatic potential $V(\vec r)$ defines  the confinement of the ballistic conductor. 
At $T=0$ the spin-dependent conductance of a mesoscopic system with two 
attached leads is given by the Landauer formula \cite{FG97}
\begin{equation} \label{LF}
G(E, B) =\frac{e^2}{h}\sum_{m', m=1}^M \ \sum_{s',s=\pm 1} |t^{m' m}_{s's}|^2 \ .
\end{equation}
The coefficient $t^{m' m}_{s's}$ denotes the transmission amplitude 
from an incoming quantum channel $m$ with spin $s$ to an outgoing channel 
with respective quantum numbers $m'$ 
and $s'$. We calculate $t^{m' m}_{s's}$ by projecting the 
corresponding Green function matrix onto the asymptotic spinors in the leads. 
As shown in Fig. \ref{ring} we define the spin directions with respect to the 
y-axis. In the case of zero flux through the ring the asymptotic spinors are 
therefore eigenstates of the Hamiltonian when entering and exiting the ring structure. In the following we focus on the case, where the two leads of width $w$ support only one open channel (M=Int$[k_F w/ \pi ]=1$).
We compute the Green function for the Hamiltonian (\ref{H}) numerically, 
using a generalized version of the recursive Green function technique based 
on a tight-binding model \cite{FG97} including spin \cite{FHR01}. 

\begin{figure}[h]
   \begin{center}      
     \psfrag{Ef}{$ \rm E_F$}
     \psfrag{Ed}{$\rm E_{kin}^\uparrow$}
     \psfrag{Eu}{$\rm E_{kin}^\downarrow$}
     \psfrag{mB}{$\rm \mu B $}

    \centering \epsfig{file=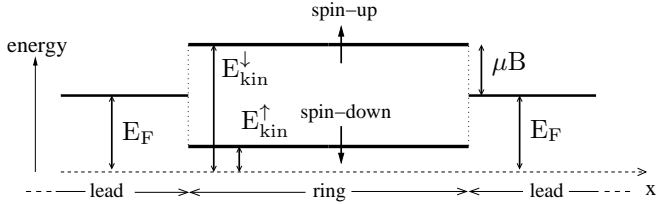, width=1.0\linewidth} 
\vspace{1mm}       
     \caption{An electron with fixed Fermi energy in the leads has different 
       kinetic energies in the ring depending on its spin state
       $ E_{kin}^{\uparrow}< E_F < E_{kin}^{\downarrow}$. 
       }
     \label{E-levels}
   \end{center}
 \end{figure}

For a  ballistic 1d ring it is feasible to obtain the eigenstates of the Hamiltonian (\ref{H}) analytically \cite{S92,Mphd}. They read $\Psi_{n,s}=\exp(i n \varphi ) \otimes \psi_n^s(\varphi)$ where the first factor desribes the motion along the ring and the second refers to the spin state $s=\uparrow , \downarrow $. The Zeeman term causes a slight difference in the kinetic energy of spin-$\uparrow$ and \mbox{spin-$\downarrow$} electrons travelling clockwise and counter-clockwise around the ring so that we must distinguish  four different orbital quantum numbers $n: n_j^\uparrow, n_j^\downarrow \ (j=1,2)$. They are given by $n'\equiv n+ \phi\  e/ h$ where the $n'$ are the solutions of the equation 
\bea \label{EV}
 \tilde E_F&=&n'^2 +n'+1/2 \\
  &\pm& \sqrt{(n'+1/2)^2 + (2n' +1) \tilde \mu B \cos\alpha + (\tilde \mu B)^2}.\nonumber
\eea
Here, $\tilde E_F=(2m^*r_0^2/ \hbar^2) E_F$ is the scaled Fermi energy, $\tilde \mu 
=(2m^*r_0^2/ \hbar^2) \mu $, and $\alpha$ the tilt angle of the textured magnetic field
with respect to the z-axis (Fig. \ref{field}). For the purpose of this paper we can specialize on very small 
homogeneous fields so that the total field is almost in-plane ($\alpha\approx \pi/2$).
In that case  the $\cos\alpha$-term under the square root is negligible,
and we get for counter-clockwise travelling waves \cite{comment-n2} the eigenvalues
\bea
{n'}_1^{\uparrow}&=&-\frac{1}{2}+\frac{1}{2} \sqrt{1+4 \tilde E - 4 
\sqrt{(\tilde \mu B)^2 + \tilde E }}, \label{n1up}\\
{n'}_1^{\downarrow}&=&-\frac{1}{2}+\frac{1}{2} \sqrt{1+4 \tilde E + 4 
\sqrt{(\tilde \mu B)^2 + \tilde E }}. \label{n1down}
\eea
The spin indices are chosen in such a way that spin-$\uparrow(\downarrow)$ labels the 
smaller (larger) quantum number $n_1$. In the adiabatic limit this corresponds to a spin 
being parallel (antiparallel) to the local field axis.
In the general,
nonadiabatic case the electron spin is not aligned with the local field direction. 
Instead, two angles $\gamma_1^\uparrow, \gamma_1^\downarrow \leq \alpha$ take 
the role of $\alpha$ and characterize the spin eigenstates which read 
\begin{equation}\label{eigenstates}
 \psi_{n_1}^{\uparrow} = \left( \begin{array}{c}
  \sin\frac{\gamma_1^{\uparrow}}{2} \\
   i e^{i\varphi}\cos\frac{\gamma_1^{\uparrow}}{2} \end{array} \right) \
   , \  \psi_{n_1}^{\downarrow } = \left( \begin{array}{c}
  \cos\frac{\gamma_1^{\downarrow}}{2} \\
  - i e^{i \varphi}\sin\frac{\gamma_1^{\downarrow}}{2} \end{array} \right)
\end{equation}
with 
\begin{equation}\label{gamma1}
  \tan\gamma_1^{\uparrow(\downarrow)} =\frac{\tilde \mu B}{  \ {n'}_1^{\uparrow(\downarrow)}+ 1/2}.
\end{equation}
In the nonadiabatic limit of a weak magnetic field the polarisation of the electron
spin remains unchanged and it is $ \gamma_1^{\uparrow(\downarrow)}\to 0$. In the 
opposite limit of strong coupling between the spin and the field the solutions
(\ref{eigenstates}) become eigenstates of the Zeeman term in the Hamiltonian (\ref{H}). 
Hence, the adiabatic limit can be defined by the requirement 
 $\gamma_1^{\uparrow(\downarrow)}\to \alpha=\pi/2$ or
 $\tan\gamma_1^{\uparrow(\downarrow)} \to \infty$. 
Expanding  Eq.~(\ref{n1up}) and (\ref{n1down}) in $\tilde \mu B 
 / \tilde E_F$  and inserting them in Eq.~(\ref{gamma1}) gives in zeroth  
order a criterion for the adiabatic regime which is independent of the spin direction:
\begin{equation}\label{Q}
 Q\equiv \frac{\tilde \mu B}{\sqrt{\tilde E_F}} \gg 1 \ .
\end{equation}
Eq.~(\ref{Q}) coincides with the condition for adiabaticity in 1d rings, which was
previously deduced by decomposing the Hamiltonian (\ref{H}) into an adiabatic and 
nonadiabatic part \cite{S92}. 
Using the first order approximation 
\be \label{approx}
{n'}_1^{\uparrow(\downarrow)}\approx -\frac{1}{2} + \sqrt{\tilde E_F} \mp \frac{1}{2}
\sqrt{Q^2+1}
\ee
we recognize that the condition for adiabaticity now becomes \emph{spin-dependent} 
and reads, for $\sqrt{Q^2 +1}\approx Q$, 
\bea
\rm{spin}-\uparrow: \qquad \frac{Q}{1- \frac{1}{2} \ \tilde \mu B / \tilde E_F} \gg 1 \ , 
\label{Qup} \\
\rm{spin}-\downarrow: \qquad \frac{Q}{1+ \frac{1}{2} \ \tilde \mu B /  \ \tilde E_F} \gg 1 \ .
\label{Qdown}
\eea
This means that spin-up electrons can reach the adiabatic regime already for lower 
magnetic fields compared to spin-down electrons. The physical origin of this behavior
can be understood as follows. In the weak adiabatic regime $Q\gtrsim 1$ the electron spin
is to a certain extent aligned with the local field direction ($0 \ll \gamma_1 < \alpha$)
and nonadiabatic spin flips are rare. Spin-$\uparrow(\downarrow)$ electrons, 
 which occupy the higher (lower) Zeeman level, have a smaller (larger) kinetic energy 
$E_{kin}^{\uparrow(\downarrow)}\approx E_F \mp \mu B$ (Fig. \ref{E-levels}). But an electron
that transverses the ring with a smaller velocity has more time to align its spin
 with the
local field direction and hence will reach the adiabatic regime more easily.
We now show how this spin-splitting in the adiabaticity parameter can lead to a 
spin polarized current in mesoscopic quasi 1d rings as illustrated in Fig. \ref{ring}. For this purpose we calculate numerically the spin-resolved transmission $T_{s,s'}=|t_{s,s'}|^2 \ (s,s'=\uparrow,\downarrow)$.
It has been shown \cite{FHR01} (see also  \mbox{Fig. \ref{switch}}) that in ballistic
symmetric ring structures supporting only one open transverse mode in the leads the
nonadiabatic components $T_{\uparrow\uparrow}$ and $T_{\downarrow\downarrow}$ vanish if
the magnetic flux is $\phi=0.5 \phi_0$ (flux quantum $\phi_0=h/e$). The remaining
adiabatic components $T_{\uparrow\downarrow}$ and $T_{\downarrow\uparrow}$, with the 
spin following the field, are  functions of the adiabaticity parameter and increase 
when increasing $Q$. 

\begin{figure}[h]
   \begin{center}      
     \psfrag{Q}{$ Q$}
     \psfrag{mB}{$\rm \mu B/E_F $}

    \centering \epsfig{file=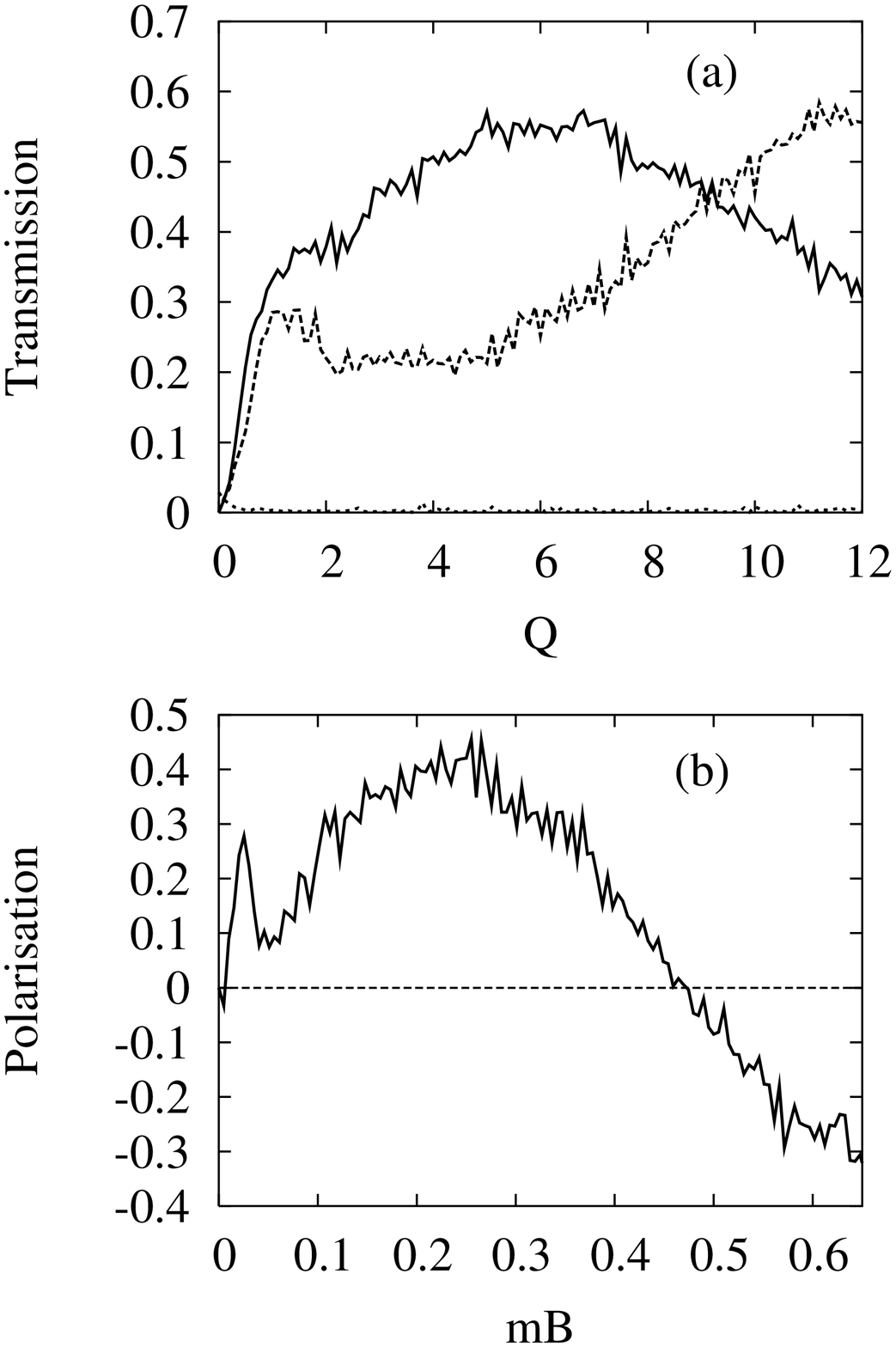, width=0.8\linewidth} 
\vspace{1mm}       
     \caption{(a) Energy averaged transmission components $T_{\downarrow\uparrow}$ (solid) and $T_{\uparrow\downarrow}$ 
       (dashed) as a function of the scaled magnetic field $Q$ for a ring with
       aspect ratio $d/r_0=0.33$, $\phi=0.5 \phi_0$ and one open channel $k_FW/ \pi =1.41-1.91$.
       The nonadiabatic components $T_{\uparrow\uparrow}$ and $T_{\downarrow\downarrow}$ 
       (dotted) almost vanish. (b) Resulting polarisation of the current as a
       function of the energy ratio $ \mu B/E_F $.}
     \label{data1}
   \end{center}
 \end{figure}

Our numerical results for the spin-resolved transmission  as a function of $Q$ for a ring
with aspect ratio $d/r_0=0.33$ and flux $\phi=0.5 \phi_0$ are depicted in Fig. \ref{data1}
(a). We have performed an energy average over half of
the first channel ($k_FW/ \pi =1.41-1.91$) to avoid strong energy dependencies due to the quantization of the angular momentum. 

The adiabatic transmission components show a clear spin splitting
$T_{\downarrow\uparrow}>T_{\uparrow\downarrow}$  for $Q \leq 9$.
As discussed before this behavior can qualitatively be understood
using Eq. (\ref{Qup}) and (\ref{Qdown}). Electrons occupying the upper Zeeman 
level travel with smaller  velocity and hence reach the adiabatic regime of maximum transmission  $T_{\downarrow\uparrow}$ at smaller $Q$-values compared to spin-down electrons which 
have transmission $T_{\uparrow\downarrow}$. This effect leads to a spin polarisation
 $ P=(T_{\downarrow\uparrow}-T_{\uparrow\downarrow})/(T_{\downarrow\uparrow}+T_{\uparrow\downarrow}+T_{\uparrow\uparrow}+T_{\downarrow\downarrow})$
of up to 45 \% (for the geometry used) 
at an energy ratio $\mu B \approx 0.25\ E_F$ (Fig. \ref{data1} (b)).
Note that in Fig 4 (b) one does observe another, sharply peaked polarisation 
maximum in the nonadiabatic regime $Q<1$, which cannot exclusively be 
attributed to the spin-dependent scaling of the adiabaticity parameter Q, as 
elucidated above. However, the numerical calculations show  a maximum value 
of 30 \% spin polarisation at a very small energy ratio $\mu B \approx 0.025\ E_F$.

Energy ratios of that order of magnitude might also become of experimental relevance.
For a 2DEG built from a GaAs/AlGaAS heterostructure with carrier densities as low as 
$n_s=1.8 \ 10^{10}\  \rm cm^{-2}$ \cite{Pf02} this energy ratio translates into a required in-plane magnetic field of $B\approx $ 500 mT. Circular inhomogeneous fields of that magnitude are well within the means of present experimental setups.
On the other hand, reaching the extended region of about 40 \% spin-polarisation (Fig. \ref{data1} (b)) would require magnetic field strengths which are an order of magnitude larger. 
This regime could be reached using diluted magnetic semiconductors (DMS) which exhibit a 
huge g-factor and therefore lead to a large Zeeman splitting \cite{CSA00}.

The depicted spin filter effect does not continue to exist for arbitrary high Zeeman splitting.  In Fig. \ref{data1} (b) we observe    a 
crossover to negative values of the polarisation at  $\mu B \approx 0.45\ E_F$, a phenomenon which becomes more pronounced with
 increasing field strength $B$. This counteracting effect results from the fact that 
for $\mu B \gtrsim E_F$  the Zeeman splitting acts as a barrier for spin-up electrons leading 
to enhanced backscattering and finally filtering of this spin state. 

We note that the nonadiabatic components $T_{\uparrow\uparrow}$ and $T_{\downarrow\downarrow}$ 
are symmetric ($T_{\uparrow\uparrow}= T_{\downarrow\downarrow}$) and do not show spin 
splitting behavior. 
In accordance with  the theoretical expectation $T_{\uparrow\uparrow}$ and 
$T_{\downarrow\downarrow}$ vanish for almost any  $Q$-value.

\begin{figure}[H]
   \begin{center}      
     \psfrag{Q}{$ Q$}
     \psfrag{mB}{$\rm \mu B/E_F $}

    \centering \epsfig{file=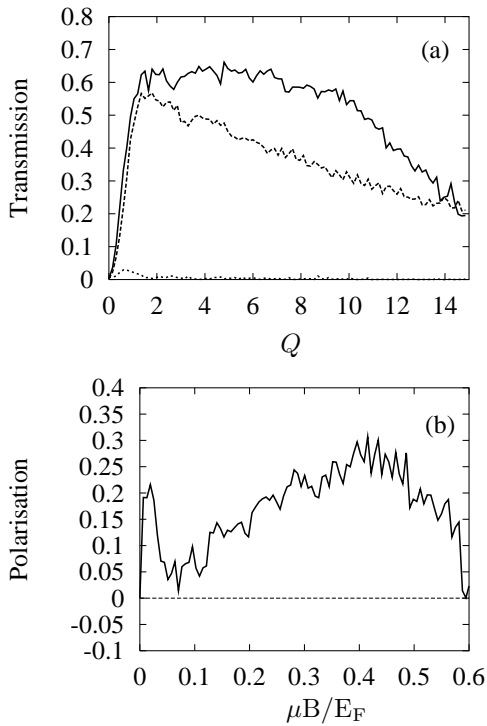, width=0.8\linewidth} 
\vspace{1mm}       
     \caption{(a) Energy averaged transmission components $T_{\downarrow\uparrow}$ (solid) and $T_{\uparrow\downarrow}$ 
       (dashed) as a function of the scaled magnetic field $Q$ for a ring with
       aspect ratio $d/r_0=0.22$. The nonadiabatic components $T_{\uparrow\uparrow}$ and $T_{\downarrow\downarrow}$ 
       (dotted)  almost vanish. (b) Resulting polarisation of the current as 
       function of the energy ratio $ \mu B/E_F $.}
     \label{data2}
   \end{center}
 \end{figure}

\begin{figure}[h]
   \begin{center}      
     \psfrag{u}{$ \uparrow  \uparrow $}
     \psfrag{d}{$ \downarrow   \uparrow $}
     \psfrag{tot}{$ \uparrow  \uparrow+ \downarrow   \uparrow $}

    \centering \epsfig{file=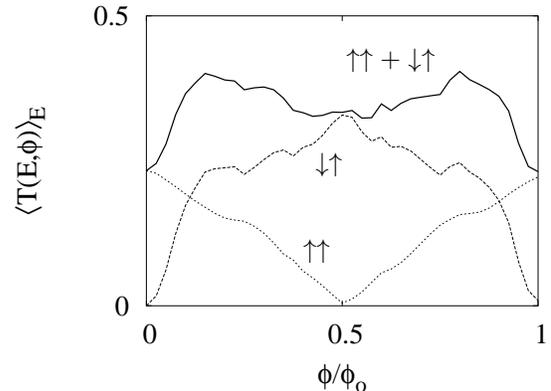, width=0.9\linewidth} 
\vspace{1mm}       
     \caption{Energy averaged transmission $T_{\downarrow\uparrow}$ 
     (dashed) and  $T_{\uparrow\uparrow}$ (dotted) for
       spin-up incoming electrons through
       a 1d ring (Fig.~\ref{ring}) as a function of magnetic flux. The 
       polarisation of the exiting current can be tuned from spin-up ($\phi =0$) to 
       spin-down ($\phi =0.5$). The total transmission 
       $T_{\downarrow\uparrow}+T_{\uparrow\uparrow}$ (solid) remains almost
       constant (from [1]). }
     \label{switch}
   \end{center}
 \end{figure}

We now study whether the spin filtering effect depends on the chosen geometry (size and aspect ratio $d/r_0$)  of the ring. Numerical results for the spin-dependent
transmission through a ring with $d/r_0=0.22$ are presented in Fig. \ref{data2} (a).
Plotting the resulting polarisation of the current (Fig. \ref{data2} (b)) we obtain a spin
filter behavior which is pretty much alike the previous one for the smaller ring. The
second polarisation maximum is now at $\mu B \approx 0.4 E_F$ and reaches only a value of
$P\approx 0.3$. The first maximum is also smaller
compared to Fig. \ref{data1} (b) but is now found at $\mu B \approx 0.01 E_F$ thus reducing the required field strength by a factor of two.  
Further numerical simulations  show that the described spin filter effect 
exists independent of the size of the ring with a maximum spin polarisation 
ranging between 30 \% and 45 \%.

\begin{figure}[h]
   \begin{center}      

    \centering \epsfig{file=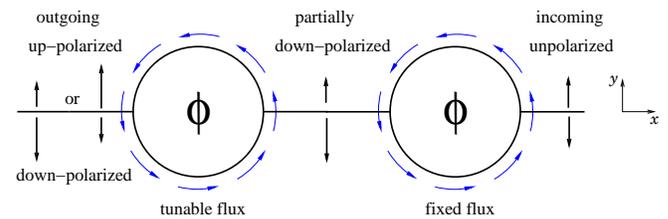, width=0.99\linewidth} 
\vspace{1mm}       
     \caption{Scheme for a serial combination of two 1d rings that could serve as
       integrated system for both creating and controlling spin-polarized 
       currents at low temperatures. The flux through the ring on the right hand side should be fixed 
       at $\phi =0.5 \phi_0$ leading to a partially down polarized current. The direction 
       of the polarisation can be changed in the second ring by applying a tunable
       flux between $\phi =0$ and $\phi =0.5 \phi_0$.}
     \label{scheme}
   \end{center}
 \end{figure}

Symmetric 1d rings do not only show a \emph{spin filter} effect for Zeeman energies $\mu B <
E_F$. Together with the \emph{spin-switch} effect described in Ref.~\cite{FHR01}
a serial connection of these  structures allows both for spin filtering and tuning of  the  polarisation of the current.  The spin-switch mechanism of Ref.~\cite{FHR01}  
is most efficient in a region of moderate magnetic fields described by $Q\approx 1$. This coincides approximately with  the $Q$ values for which the first polarisation maximum of up to 30 \% arises. As illustrated in Fig. \ref{switch} for spin-up incoming electrons, applying zero flux leaves the spin polarisation unchanged. For half a flux quantum   the polarisation is reversed and the electron exits in a spin-down state. Analogous results hold for spin-down incoming electrons. 
In Fig. \ref{scheme} we propose a configuration of two ballistic 1d rings as in Fig. \ref{ring}  that could serve as a device to coherently create and control spin-polarized currents. The inhomogeneous magnetic field in both rings can be chosen approximately equal and in such a way that $Q\approx 1$. Applying a fixed flux $\phi =0.5 \phi_0$ to the  ring on the right hand side will create a partially spin-down polarized current. The flux through the second ring is tunable between $\phi =0$ and $\phi =0.5 \phi_0$ and controls the direction of the polarisation. \\
We note that an analogous scheme could also be realized using radial inhomogeneous magnetic fields \mbox{($\vec B \sim \hat r$ )} instead of circular ones. Such fields can be created by placing a micromagnet in the center of the disk \cite{YTW98}. In that case the polarisation of the current would be defined with respect to the x-axis. 

However, we do not expect the described spin filter effect to exist for inhomogeneous effective fields resulting from a strong spin-orbit Rashba coupling \cite{Yau}. The Rashba effect leads to a spin splitting in the dispersion relation but electrons occupying the two spin branches travel with equal group velocity and hence have the same adiabaticity parameter $Q$.

Finally we summarize the conditions which have to be met to make the scheme of 
Fig.~\ref{scheme} work.  Both the filter and switching mechanism are based on coherent transport and hence exist only at low temperatures. The ballistic ring structures have to be symmetric with respect to the x-axis and may only support one open channel in the leads. There is no restriction on the number of transverse modes in the ring arms themselves. \\
Although our proposed setup is based on  a ballistic system we  have numerical 
evidence that the observed spin filter effect also exists in disordered 
rings, however, it is less pronounced. Since the random distribution of the 
impurity potential breaks the symmetry of the ring, the nonadiabatic 
transmission components no longer vanish independently of the value of the 
adiabaticity parameter $Q$ . Especially for $Q<1$ they give a strong contribution 
to the transmission and thus decrease the polarisation considerably. Given 
that the mean free path $\ell$ depends on the (spin-dependent) travelling 
velocity of the electrons (weak scattering limit)  the maximum value of the 
spin polarisation in the adiabatic limit is also lowered. The reason is that 
the adiabaticity parameter in the diffusive regime scales with the inverse 
square root of the mean number of scattering events ($Q / \sqrt{N} \sim Q \  \ell / L$) \cite{PFRunp} which counteracts the spin-dependent scaling  of Q described above.

To summarize we have studied the spin-dependent transport properties of ballistic symmetric 1d ring structures subject to an inhomogeneous magnetic field. Our numerical calculations show that a moderate Zeeman splitting compared to the Fermi energy can lead to an enhanced transmission probability for electrons occupying the \emph{upper} Zeeman level, which results in a spin polarization of the current of up to 45\%. This effect exists in the adiabatic regime and can qualitatively be understood in terms of a spin-dependent adiabaticity parameter $Q$. Further increase of the strength of the inhomogeneous field leads finally to a reversal of the effect, because it is counteracted by the usual spin filtering of the charge carriers at a large Zeeman barrier.\\
 For very small Zeeman spitting our numerical calculations  indicate  the existence of another, sharply peaked  polarisation maximum of up to 30 \%. This  effect, which is not yet fully understood, could be observed for a strength of the inhomogeneous field that is well within the means of present experimental setups. \\
Moreover we make use of the predicted spin-switch effect in such a system and propose an integrated device  consisting of two rings connected in series that in principle can serve both as spin filter and spin controller for currents  at low temperatures.

%
%


\begin{thebibliography}{10}
\bibitem{comment-n2}
For the sake of simplicity we restrict the discussion  on counter-clockwise travelling
waves, but it is straightforward to show that identical results  must hold for  
clockwise travelling waves.
\end{thebibliography}



\end{document}